\documentstyle[prl,aps,multicol,overcite]{revtex}
\begin{document}
\title{Tests of Bell Inequalities}

\author{Lev Vaidman}

\maketitle
\vspace{.4cm}
 
\centerline{  Centre for Quantum Computation}
\centerline{ Department of Physics, University of Oxford,}
\centerline{ Clarendon Laboratory, Parks Road, Oxford OX1 3PU, England.}
 \vskip .2cm
\centerline{and}
 \vskip .2cm
\centerline{ School of Physics and Astronomy}
\centerline{Raymond and Beverly Sackler Faculty of Exact Sciences}
\centerline{Tel-Aviv University, Tel-Aviv 69978, Israel}

\date{}

\vspace{.2cm}
\begin{abstract}
  According to recent reports, the last loopholes in testing Bell's
  inequality are closed. It is argued that the really important task
  in this field has not been tackled yet and that the leading
  experiments claiming to close locality and detection efficiency
  loopholes, although making  very significant progress, have
  conceptual drawbacks. The important task is constructing quantum
  devices which will allow winning games of certain correlated replies 
  against any classical team. A novel game of this type is proposed.
 \end{abstract}

\vskip .2cm
\begin{multicols}{2}

Quantum mechanics predicts unusual correlations between outcomes of
particular experiments in space-like separated regions. The
peculiarity of these correlations is that they are stronger than any
correlations explainable by a local theory. The quantum correlations,
as was proven by Bell \cite{Bell}, break certain inequalities which
have to be fulfilled if the results of every experiment are determined
by some local hidden variable (LHV) theory.  Recently, we have been
witnessed an outstanding progress in the tests of Bell's inequalities,
but a decisive experiment which would rule out any LHV theory has not
been performed yet. The deficiencies of the experiments are described
by {\it locality} and {\it detection efficiency} loopholes \cite{Pe}.
According to recent reports, Weihs {\it et al.}  \cite{nolo} closed
the locality loophole and Rowe {\it et al.} \cite{delo} closed the
detection efficiency loophole.
I want to point out that the leading experiments claiming to
close the loopholes \cite{nolo,delo}, although clearly making very
significant progress, have conceptual drawbacks. These drawbacks have
to be removed before claims like ``the last loophole closes'' can be
made.

 Moreover, I want to argue that the experimental efforts should be
turned to a slightly different task: instead of showing that the
quantum correlations cannot be explained by local theories, it is more
important to show that the quantum correlations can be used.

Today, there is a firm consensus that there is no real question what
will be the outcome of this type of experiment: the predictions of
quantum theory or results conforming with the Bell inequalities.
Predictions of quantum mechanics have been verified in so many experiments
and with such unprecedented precision that, in spite of the very
peculiar and nonintuitive features that Bell-inequality experiments
demonstrate, only a minute minority of physicists believe that quantum
mechanics might fail in this type of experiment.  However, the fact
that we are pretty sure about the final result of these experiments
does not mean that we should not perform them. One goal of such
experiments is to change our intuition which developed from observing
classical phenomena. But more importantly, these experiments should
lead to the stage in which we will be able to {\it use} these unusual
correlations.

Conceptually, the most simple, surprising, and convincing out of the
Bell-type experiments is the Mermin's version of the
Greenberger-Horne-Zeilinger (GHZ) setup \cite{Mer}.  I find that it
can be best explained as a game \cite{GHZV}. The team of three players
is allowed to make any preparations before the players are taken to
three remote locations. Then, at a certain time, each player is asked
one of two possible questions: ``What is $X$?''  or ``What is $Y$?''
to which they must quickly give one of the answers: ``$1$'' or
``$-1$''.  According to the rules of the game, either all players are
asked the $X$ question, or only one player is asked the $X$ question
and the other two are asked the $Y$ question. The team wins if the
product of their three answers is $-1$ in the case of three $X$
questions and is $1$ in the case of one $X$ and two $Y$ questions.  It
is a simple exercise to prove that if the answer of each player is
determined by some LHV theory, then the best strategy of the team will
lead to 75\% probability to win. However, a quantum team equipped with
ideal devices can win with certainty. Each player performs
 a spin measurement of a spin-1/2 particle:
$\sigma_x$ measurement for the $X$ question and $\sigma_y$ measurement
for the $Y$ question and gives the answer $1$ for spin ``up'' and $-1$ 
for spin ``down''. Quantum theory ensures that if the players have
particles prepared in the GHZ state, the team always wins.
Actually constructing such devices and seeing that,
indeed, the quantum team wins the game with probability significantly
larger than 75\% will be a very convincing proof of Bell-type
inequalities. 

The game need not be based on the GHZ-type ``Bell
inequality without inequalities'' proof. 
A two-party game based on another proof is presented at the end of
 the letter.

A game-type experiment, if successful, will definitely close the detection
efficiency loophole. If it is also arranged that the party asking the
questions chooses them ``randomly'' (more about randomness below),
then it will also close the locality loophole. Note, that an
experiment which simultaneously closes both the detection efficiency
and the locality loophole will not necessarily be suitable for winning
games of the type I described here. 
 Indeed,
the current GHZ experiment \cite{GHZe}, even if performed with ideal
optical devices, cannot help to win the GHZ game. At no stage of this
experiment is there a pure GHZ state. Only when the three photons were
detected at three different detectors in coincidence with the fourth
(trigger) photon, could we claim that the polarization state of the{\it detected} photons was the GHZ state. If a player in the team
detects the trigger photon and makes a polarization measurement (the
analog of the spin measurement) according to the question he is asked, he
cannot be sure that the team will win since the other players would know
the good answers only if they detect the photons too. However, the
setup is such that there is a high probability for this not to happen
even if they have 100\% efficient detectors.

Let me
now discuss the current experiments.  In order to close the locality
loophole Weihs {\it et al.}\cite{nolo} used fast quantum experiments to
choose between local measurements.  The results of these experiments
are ``genuinely random'' according to the standard quantum theory, but
are {\it not random} in the framework of LHV theories.  Their outcomes
are also governed by some LHVs in each site. There is enough time for
information about these LHVs to reach other sites before the
measurements there took place and, therefore, the locality loophole is
not closed.  The experiment, nevertheless, is a significant step
forward because its results can be explained only by a higher level
LHV theory in which hidden variables specifying the behavior of one
system are influenced by hidden variables of other systems.

A frequently discussed experiment in which a person at each site will
have enough time to exercise his ``free will'' to choose between the
measurements will be very convincing, but conceptually, not much
better: we cannot rule out the existence of LHVs which are responsible
for our seemingly ``free'' decisions.  A better experiment for closing
the locality loophole is to make the choice of the local measurements
dependent on the detection of photons arriving from space-like
separated events in distant galaxies.  Then, the LHV-type explanation will
be a ``conspiracy'' theory on the intergalactic scale.

Now I will discuss the latest experiment by Rowe {\it et al.}
\cite{delo} who claimed to close the detection efficiency loophole. In
this experiment the quantum correlations were observed between results
of measurements performed on two ions few micrometers apart.  The
detection efficiency was very high. It was admitted that the locality
loophole was not closed, but the situation was worse than that.
Contrary to other experiments \cite{nolo}, not only the information
about the choice, but also about the {\it results} of local
measurements could reach other sites before completion of measurements
there.  The reading of the results was based on observing numerous
photons emitted by the ions. This process takes time which is a few
orders magnitude larger than the time it takes for the light to go
from one ion to the other. Thus, one can construct a very simple LHV
theory which arranges quantum correlations by ``communication''
between the ions during the process of measurement. It is much simpler
to construct a LHV theory which employs also ``outcome dependence''
instead of only ``parameter dependence'' \cite{Ja}.

The purpose of closing the detection efficiency loophole was to rule
out the set of LHV theories in which the particle carries, among
others, instructions of the type: ``if the measuring device has
particular parameters, do not be detected''. Such hidden variables
cannot explain the correlations of the Rowe {\it et al.} experiment
and this is an important achievement.  However, the task of performing
an experiment closing the detection efficiency loophole without
opening new loopholes (the possibility for ``outcome dependence'' LHV
in Rowe {\it et al.}) is still open.

Recently the bizarre features of quantum mechanics have been explained
through various games. Apart from the GHZ game described above there
have been several other proposals: an interesting variation of the GHZ game by Steane
and van Dam \cite{SvD}, a game based on the original Bell proof by
Tsirelson \cite{Tsi}, the ``quantum cakes'' game based on a
non-maximally entangled state by Kwiat and Hardy \cite{KH} (see
related experiment \cite{BGNP}).
 Note also the proposal of Cabello \cite{Ca} for a two-party
Bell-inequality proof which can be transformed into a game too.  Let
me present here one more game. My game is called an ``impossible
necklace'' and it is based on the Zeno-type Bell inequalities proof
\cite{SHB}.

A team of two players wants to persuade a third party, ``the
interrogator,'' that they found a secret of making an ``impossible
necklace''. The impossible two-colored neclace has an even number of
beads $N$ and all adjacent beads are of different colors except beads
1 and $N$ which are of the same color.  The team does not want to
reveal the ``secret coloring'', but the players are ready to reveal
the colors of any two adjacent beads of the necklace. They claim to
have identical necklaces of this kind, one necklace for each player.
The interrogator arranges to ask one player the color of any single
bead and ask the other player, at a space-like separated region, the
color of one of the adjacent beads.  If the team succeeds in giving
the correct answers in many repeated experiments (with new necklaces
each time), a naive interrogator might be persuaded that the team
knows how to make such necklaces.  Indeed, if it is a ``classical
team'', and the players decide in advance what answer they will give
for every question, then the probability to fail is at least $1/N$.
(There are $N$ different pairs and there is no way to arrange that all
have correct coloring.)  Therefore, the probability to pass the test,
say $5N$ times is
  \begin{equation}
    \label{clas}
 prob_{\rm classical} =(1-{1\over N})^{5N} \sim e^{-5} \sim 0.01 .   
  \end{equation}
The quantum team can do much better. The players do not make any
necklaces. Each player  take with him
a
spin$-{1\over 2}$ particle from the EPR pair. When a player is asked
the color of a bead $i$, he measures spin component in the direction
 $\hat\theta_i$  in the $x-z$ plane which  makes an
angle $\theta_i = {{\pi i}\over N}$ with the $z$ axes. He says
``green'' if the result is ``up'', and ``red'' if the result is
``down''.  His partner do the same. For all pairs,
the measurements are in the directions which differ by the angle
${{\pi }\over N}$  except for the pair $\{1, N\}$ in which case the
angle is ${{\pi  (N-1)}\over N}$. Therefore, the probability to fail
the test is   $\sin^2 {{\pi }\over {2N}}$.
 The
probability to pass $5N$ tests is
 \begin{equation}
    \label{qua}
prob_{\rm quantum} = (1 - \sin^2 {{\pi }\over {2N}})^{5N} \sim
(1-{\pi^2\over {4N^2}})^{5N} \sim e^{{-5\pi^2\over {4N}}} .
\end{equation}
For $N=100$ the quantum team has  probability of almost 90\% to succeed, compare with 1\% of a classical team. 

Technological problems will not allow an experiment with a large
number $N$ in a near future. Putting aside the attempt to ``fool'' the
interrogator that the team has impossible necklaces, the game can be
defined as the competition of two-player teams to pass the interrogator
tests a maximal number of times. For any number $N \geq 4$, the quantum
team has an advantage over a classical team, so this game is a
realistic proposal for demonstrating Bell-type
inequalities. (Certainly more realistic than the GHZ game which
requires a three-particle source.)

The ideal situation (in the sense that it closes all loopholes) is
that the questions which players are asked are decided by fast
detectors obtaining signals from space-like separated distant galaxies
and that the players give their answers quickly enough such that the
communication between them is impossible. However, I do not think that
the stringent requirement of closing the locality loophole is very
important. It seems to me that an implementation of one of these games
with players placed in separate sealed rooms which prevent them from
sending out any signals will be a very dramatic and persuasive proof
that the nature indeed has the peculiar features predicted by quantum
theory. More importantly, it will show that quantum technology is
capable of performing communication tasks which are impossible when
classical devices are used.

 It is a pleasure to thank  Jonathan Jones,  Lucien Hardy,  and Hans Christian von Bayer for helpful discussions.
 This research was supported in part by grant 471/98 of the Basic
 Research Foundation (administrated by the Israel Academy of Sciences
 and Humanities) and the EPSRC grant  GR/N33058.

\end{multicols}

\begin{thebibliography}{99} 
\footnotesize

\bibitem{Bell}
J.S. Bell,
 Physics  1 195 (1964).

\bibitem{Pe}
P. Pearle, 
Phys. Rev.  D 2, 1418  (1970).

\bibitem{nolo}
G. Weihs  et al.,
 Phys. Rev. Lett.  81, 5039 (1998).
 
\bibitem{delo}
M.A. Rowe,   et al., Nature  409, 791 (2001).

\bibitem{Mer}
D. N. Mermin,
 Am. Jour.  Phys.  58, 731 (1990).

\bibitem{GHZV}
L. Vaidman,
 Found. Phys.,  29, 615 (1999).

\bibitem{GHZe}
J.W. Pan  et al.,
  Nature  403, 515  (2000).
 
\bibitem{Ja}
J. P. Jarrett, Nous 18, (1984).

\bibitem{SvD}
A.   Steane and W. van Dam, 
Phys. Today.  53, 35 (2000).

\bibitem{Tsi}
B. Tsirelson,
Lecture Notes, Tel-Aviv University (1996).

\bibitem{KH}
P.G. Kwiat and L. Hardy,
 Am. Jour.  Phys.  68, 33 (2000).

\bibitem{BGNP}
G. Brida et al., Phys. Lett. A 268 (2000).

\bibitem{SHB}
E.J. Squires, L. Hardy, and H.R.  Brown,  Stud. Hist. Phil. Sci.  25, 425 (1994).

\bibitem{Ca}
A. Cabello
Phys. Rev. Lett. 86, 1911 (2001).
\end{thebibliography}
\end{document}